\documentclass[a4paper]{article}
\usepackage[latin1]{inputenc} %
\usepackage[T1]{fontenc} %
\usepackage{RR}
\usepackage{hyperref}
\RRdate{January 2010}

\RRauthor{
Manos Dramitinos%
  \and
Isabelle Gu\'erin Lassous%
}
\authorhead{Dramitinos \& Gu\'erin}
\RRtitle{M\'ecanismes d'allocation de bande passante bas\'es sur les ench\`eres pour les r\'eseaux sans fil du futur}
\RRetitle{Auction-based Bandwidth Allocation Mechanisms for Wireless Future Internet}
\titlehead{Auction-based Bandwidth Allocation Mechanisms}
\RRnote{This work was partially supported by the
    European integrated project AEOLUS.}
\RRnote{mdramit@gmail.com - Isabelle.Guerin-Lassous@ens-lyon.fr}
\RRresume{Un aspect cl\'e de l'Internet du futur sera la gestion efficace des ressources et sp\'ecifiquement des ressources sans fil. Nous \'evoluons actuellement vers des r\'eseaux d'acc\`es sans fil int\'egr\'es qui doivent permettre la mobilit\'e des utilisateurs, un acc\`es transparent et une certaine qualit\'e de service. Le terme "int\'egr\'e signifie que les derniers sauts sans fil de l'Internet du futur comprendront plusieurs r\'eseaux sans fil ayant des capacit\'es et des zones de couverture h\'et\'erog\`enes. La gestion efficace des ressources de ces r\'eseaux d'acc\`es sans fil est cruciale \'etant donn\'ee la limitation des ressources radios. Dans cet article, nous proposons un m\'ecanisme bas\'e sur la notion d'ench\`eres pour allouer de la bande passante dans de tels r\'eseaux. Avec ce m\'ecanisme, l'efficacit\'e est atteinte, i.e. le bien-\^etre social est maximis\'e. Nous d\'efinissons une suite de tels jeux afin de prendre en compte les probl\`emes d'utilit\'e des utilisateurs et d'incitation. Enfin, nous \'etendons ce m\'ecanisme aux sessions multicast. Nous analysons aussi la complexit\'e en calcul et les surco\^ut en terme de messages des m\'ecanismes propos\'es. Enfin, nous montrons comment transformer l'ench\`ere en un m\'ecanisme coop\'eratif capable de prioritiser certaines classes de services en rempla\c cant les paris des utilisateurs par des poids g\'en\'er\'es par le r\'eseau. Pour finir, des simulations de ces m\'ecanismes sont propos\'ees.    
}
\RRabstract{An important aspect of the Future Internet is the efficient utilization of
(wireless) network resources. In order for the - demanding in terms of QoS
- Future Internet services to be provided, the current trend is evolving 
towards an ``integrated'' wireless network access model that enables 
users to enjoy mobility, seamless access and high 
quality of service in an all-IP network on an ``Anytime, Anywhere'' basis.
The term ``integrated'' is used to denote that the Future Internet wireless 
``last mile'' is expected to comprise multiple heterogeneous geographically 
coexisting wireless networks, each having different capacity and coverage 
radius. The efficient management of the wireless access network resources is
crucial due to their scarcity that renders wireless access a potential
bottleneck for the provision of high quality services.
In this paper we propose an auction mechanism for allocating the 
bandwidth of such a network  
so that efficiency is attained, i.e. social welfare is maximized. 
In particular, we propose an incentive-compatible, efficient 
auction-based mechanism of low computational complexity. 
We define a repeated game to address user utilities and incentives issues.
Subsequently,  we extend this mechanism so that it can also 
accommodate multicast sessions.
We also analyze the computational complexity and message 
overhead of the proposed mechanism.
We then show how user bids can be replaced from weights generated by the
network and transform the auction to a cooperative mechanism capable of
prioritizing certain classes of services and emulating DiffServ and time-of-day
pricing schemes.  The theoretical analysis
is complemented by simulations that assess the proposed mechanisms properties
and performance. We finally provide some concluding 
remarks and directions for future research.}
\RRmotcle{r\'seaux sans fil h\'et\'erog\`enes, allocation de bande passante, ench\`eresc}
\RRkeyword{Heterogeneous wireless networks, bandwidth allocation, auction theory}
\RRprojets{Reso}
\RRdomaine{D3} 
\RRtheme{R\'eseaux et T\'el\'ecommunications}
\URRhoneAlpes 
\RCGrenoble 

\begin{document}
\RRNo{7188}
\makeRR   

\section{Introduction}\label{sec:intro}

An important aspect of the Future Internet is the efficient utilization of
(wireless) network resources. In order for the - demanding in terms of QoS
- Future Internet services to be provided, the current trend is evolving 
towards an ``integrated'' wireless network access model that enables 
users to enjoy mobility, seamless access and high 
quality of service in an all-IP network on an ``Anytime, Anywhere'' basis.
The term ``integrated'' is used to denote that the Future Internet wireless 
``last mile'' is expected to comprise multiple heterogeneous geographically 
coexisting wireless networks, each having different capacity and coverage 
radius \cite{thesis}. This integrated wireless network access approach has been 
also motivated by the 3G cellular mobile networks and their potential enhancement 
with WLAN radio access, 
and also WiMax,  which can  
utilize the analogue TV bands (700 MHz) that will be made available 
with the upcoming roll out of digital TV. 
Thus, integrated wireless network access can comprise the 
means of providing wireless Future Internet services to possibly mobile users.

\begin{figure}[htbp]
	\centering\includegraphics[width=.6\columnwidth]{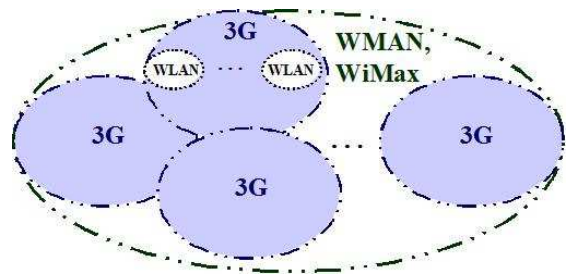}
	\caption{Co-existing wireless access networks.}
	\label{fig:4g}
\end{figure}

In this paper we propose, analyze and assess an auction-based mechanism 
for allocating the downlink bandwidth 
of such a Wireless Future Internet network whose access architecture is 
hierarchical in terms of radius and thus geographical coverage. 
In particular, we consider a single network
provider who owns and thus controls such a wireless access infrastructure,
as the one depicted in Fig.~\ref{fig:4g};  
we address the problem of deciding on which user flows to admit  
and how the downlink bandwidth of the ``integrated'' access network
should be allocated to the competing user services 
so that {\em efficiency} is attained, i.e. {\em social welfare} is maximized. 
Social welfare is defined as the sum of all users utilities and is a widely
used and commonly accepted maximization goal which depicts - as its name indicates - 
the society's attained welfare from the adoption of a certain scheme.

We propose an auction mechanism, applicable in Wireless Future Internet 
hierarchical networks, comprising of multiple tiers of wide, middle and 
local area access networks owned by one operator. 
Due to the hierarchical structure of the network, depicted as
Fig.~\ref{fig:4g}, each user can be served by means of either a higher 
tier network of that area, or by some lower tier network (e.g. a WLAN). 
The standard assumption that user terminals
are capable of accessing multiple network interfaces is made. Our model
is motivated by the 4G integrated access model and the
fact that wireless operators invest in WLANs as a cheaper means
of providing high speed download services compared to other technologies 
such as HSDPA \cite{epress}.
Finally, terminals capable of connecting to multiple network 
interfaces are available in the market \cite{nokia}.  

We then extend the proposed mechanism so as to
support {\em multicast}. Multicast is a very promising technology that
enables the efficient transmission of only one copy of
data to a group of receivers. This is very useful
for the wireless networks where due to the exponentially increasing
demand for high-speed Future Internet data services and the scarcity 
of the wireless spectrum, it is imperative to utilize the network efficiently.

The remainder of this paper is organized as follows: Section~\ref{sec:rwork} 
contains an overview of related work. In Section \ref{sec:auc}
we present the proposed auction mechanism and its properties. We  
extend our mechanism in Section~\ref{sec:mc} so as to support 
multicast. Section~\ref{sec:time} studies utility definition 
and user incentives issues
in a repeated game comprising of consecutive auctions. In Section~\ref{sec:coop} we 
transform the auction mechanism to a cooperative  
mechanism capable of prioritizing services and emulating 
DiffServ and time-of-day pricing schemes. Section~\ref{sec:assessment}
assesses our mechanism's computational complexity. 
Section~\ref{sec:sims} complements the analysis of the proposed
mechanism by means of simulations.
Finally, Section~\ref{sec:concl} provides some concluding remarks and interesting 
directions for future research.

\section{Related Work}\label{sec:rwork}

A plethora of architectures, protocols, resource management and handoff schemes 
for 4G and Wireless Future Internet have been proposed in the 
literature (see \cite{thesis}, \cite{merakos}, \cite{arch}, \cite{sas}, \cite{dm}, 
\cite{handoff} and references therein). Those proposals generally lack 
economic merit, since they do not prioritize users in terms of 
their utility for the service, as opposed to our approach. In 
fact, most of them are complements to our scheme, since they provide 
the technological solutions by means of which our scheme can be 
applied. Thus, we henceforth restrict attention to similar economic-aware schemes. 

Hierarchical
bandwidth allocation has been studied in \cite{marina}, where in the top tier 
a unique seller allocates bandwidth to intermediate providers (e.g. Internet 
Service Providers), who in turn resale their assigned shares of bandwidth to 
their own end customers in the lowest tier. This model involves
resale among the 3 tiers pertaining to different actors, as opposed to
our case where a single actor owns {\em multiple} tiers and aims to efficiently 
allocate their bandwidth.

A utility-based load balancing scheme in WLAN/UMTS networks is proposed in \cite{umts}. 
For each network, a utility reflecting the current load is computed. The values
of the network utilities, i.e. loads, are communicated to the clients who can switch to
the less loaded  network. Thus, this scheme cannot prioritize users in terms of their
utilities. 

Closest related to our work is the scheme of \cite{gds}, where {\em fixed rate pipes}
over {\em two} alternative paths are auctioned among synchronized users having different 
utilities for the two paths and thus submitting different bids. The latter
contradicts the seamless access concept, as opposed to our work.
Also, the scheme of \cite{gds} cannot be generalized 
for multiple services, rates and networks, as opposed to our scheme.
Finally, the computational complexity of the mechanism of \cite{gds} exceeds 
that of our scheme, mostly due to the fact that different utility values per user
are assigned to each path. 

\section{The Proposed Auction}\label{sec:auc}

Users compete for downlink data services, such as FTP, 
video and audio streaming over a Future Wireless Internet access network 
depicted as Fig. \ref{fig:4g}. 
Each service flow is shaped by the network operator
in a similar way with the 3G networks, i.e. by means of token buckets 
\cite{3gpp}. The use of token buckets allows an accurate description - and
shaping/policing - of the load injected by the various flows over the network and
thus a precise estimate of the multiplexing capabilities of the network
under various loads. 
Therefore, in our model different services $s_i, s_j \in \mathcal{S}$ may
differ only in terms of their respective mean rates $m_i \neq m_j$. 

We propose a sealed-bid auction, run 
periodically for allocating bandwidth over a given period of time, i.e. users
are synchronized. Dynamic user arrivals and departures may occur at the various 
auctions; this is discussed in Section~\ref{sec:time}. 

Each user $i \in \mathcal{I}$ has a certain utility $u_i$ 
and declares a willingness to pay $w_i$ for a service of rate $m_i$, by 
submitting $w_i$ as part of his service request. Let $p_i$ denote the per unit 
of bandwidth willingness to pay of user $i$, i.e. $p_i = w_i / m_i$.

For an $L-$tier network architecture, let $C_k^{(l)}$ denote the capacity of 
the $l$-tier access network $k$, with $l = {1,..,L}$; i.e. $l=1$
corresponds to the network technology having the greatest geographical coverage. 
$k$ is the index of the $l$-tier network accessible by the user, e.g. the 
index/ESSID of a WLAN inside the coverage of which the user is located. 

Users are both unaware of and unable to control the internal routing of 
their traffic. However, since assigning a possibly mobile user to a network 
interface of low geographical coverage is expected to result to higher number 
of handoffs for that user over time, our scheme
attempts to assign high value users to the network interface with the highest radius.
Thus, the declaration of a high willingness to pay from a user also implicitly results 
in a lower expected number of handoffs. This comprises an additional attractive feature of 
the proposed mechanism.

\begin{figure}[htbp]
	\centering\includegraphics[width=.6\columnwidth]{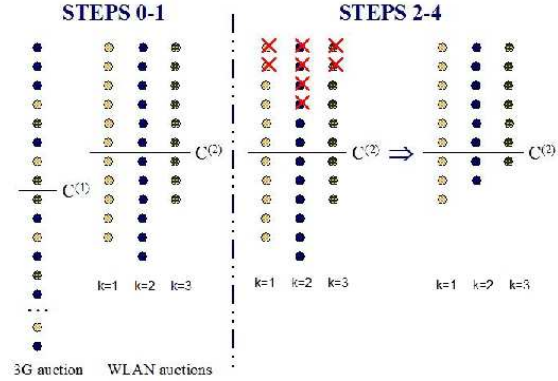}
	\caption{Sample algorithm execution for a UMTS/WLAN network. Different
	colors denote the different WLANs that users can utilize, depending on their location.}
	\label{fig:algo}
\end{figure}

Upon a service request is received, the operator\footnote{The terms network and operator
are used interchangeably in the remainder of the paper to refer to the owner of the network, who is also the auctioneer.} creates the bids 
$b_i^{(l)} = (p_i, m_i)$ $\forall i \in \mathcal{I}$, $l=1,..,L$ and updates the 
respective active bids sets $\mathcal{B}_k^{(l)}$ for all networks $k$
accessible by the user, one per tier. 
The basic idea is that winner determination is performed starting from
the highest coverage network, where competition is most fierce. The 
users bids are sorted by $p_i$ and given the capacity constraint, the 
highest of them are declared as winning. The auction winners are propagated to the
lower tier network auctions, from which their bids are deleted. Winner determination is 
then performed for the next tier, until the lowest tier is reached; this is done
simultaneously, in a distributed fashion for same-tier networks.
A sample auction execution for a two-tier network comprising of a 3G network
and three WLANs is provided as Fig.~\ref{fig:algo}; in order to keep the
presentation of the auction simple,
there is no
top-tier WiMax or LTE interface in this simplified example.

The proposed auction is defined as follows:
\subsubsection*{Step 0} Set $l = 1$. Sort($\mathcal{B}_k^{(l)}$) $\forall k, l$ // sort bids per $p_i$.
\subsubsection*{Step 1} Determine winning bids $\mathcal{W}_k^{(l)}$ of the $l$-tier network $k$ 
to be the {\em largest} set of the {\em highest} bids of $\mathcal{B}_k^{(l)}$ that do not violate 
the capacity constraint $C_k^{(l)}$.
\subsubsection*{Step 2} For every user $i$ having bid $b_i \in \mathcal{W}_k^{(i)}$ delete 
user $i$'s bids from $\mathcal{B}_k^{(j)}$, $\forall j > l$.  
Set $l = l + 1$.
\subsubsection*{Step 3} If $(l < L)$ goto {\em Step 1}. 
\subsubsection*{Step 4} Compute payments.

\subsection{Incentive compatibility and efficiency}
The payment rule of the auction (Step 4) should enforce truthful bidding,
i.e. $w_i = u_i \  \forall i \in \mathcal{I}$, so that the available network resources
can be assigned to the users that value them the most and thus social welfare is maximized. 
A strong result of auction theory is that the {\em Vickrey-Clarke-Groves (VCG) mechanism}, 
is the essentially unique mechanism where it is dominant strategy for the bidders to bid 
truthfully, the outcome maximizes social welfare and bidders having zero valuations 
attain zero benefit \cite{krishna}. Note that the term ``essentially unique'' implies that any 
auction mechanism that has the same outcome with this generalized Vickrey auction is
essentially an equivalent specification of the VCG mechanism.

This mandates that besides awarding the items auctioned to the highest positive bids
submitted by the users, we need to apply the VCG payment rule. The
latter defines each user's charge to be the social opportunity cost that his 
presence entails. Formally, user $i$ is charged:

\begin{equation}
\label{eq:vcg}
SW_{-i}(0, \theta_{-i}) - SW_{-i}(\theta)
\end{equation}

$SW_{-i}$ denotes the social welfare of bidders other than $i$, $\theta$ is the set of the
users' reported valuations and $(0, \theta_{-i})$ is the efficient outcome
if $i$'s reported value were $0$ and the other users' reports remain
unchanged. More intuitively, each user's charge equals the
losing bids that would be winning in the auction if his own bid were set to 0. 
This amount is both unaffected by and less than the user's own bid.
In general, this rule requires that in order to compute the charge of every winner
in the auction, the auction must  be rerun by removing this user from the auction,
so that his charge can be computed from \ref{eq:vcg}. This is a tedious procedure,
resulting in the general case in NP computational complexity. 

However, our mechanism takes advantage of its hierarchical structure:  
Note that all user bids are propagated to the upper tiers so that
winner determination is performed and then the winners are propagated downstream
so that the auction proceeds at the lower tiers. A similar upstream update
can be made for the users that will win in the lower tiers. That is, 
the winning bids of the  $l+1$,..,$L$-tier auctions are deleted from
the local losing bids index of a $l$-tier auction. Thus, the information 
required to determine the 
``global'' social opportunity cost is available {\em locally} per auction.
Hence, each winner's $i$ charge is computed as the sum of the highest (locally
stored) losing bids whose sum of rates equals $m_i$ and 
there is no need to rerun the auction; we elaborate more on this
in Section~\ref{sec:assessment}.

{\bf Proposition 1:} The proposed auction is efficient.
\\
\noindent{\bf Proof:\,}
By construction, our auction examines all the bids
at the top tier and admits the highest. This is repeated for all tiers,
making impossible not to admit a bid that is higher than those admitted. 
Since users bid truthfully due to the VCG payment rule, i.e. $w_i = u_i$,
and the highest bids are admitted, social welfare is maximized. Thus, 
efficiency is attained,  and after the algorithm terminates it is impossible for
a winning bid to be lower than a losing bid.
\\ 

\subsection{Revenue} 
Since our mechanism is essentially a VCG auction, it attains 
the highest revenue among {\em all}
efficient mechanisms \cite{krishna}. 

\section{Supporting Multicast}\label{sec:mc}

In this section, we extend the auction of Section~\ref{sec:auc} so as
to support multicast. In particular, our auction is complemented
with the operator's decision on whether a user is served by means of  
unicast or multicast. Hence, this is decided by the operator 
as a network optimization decision, opaque to the users.
Thus, it is not part of the
user's strategy space to choose between a unicast or a multicast service session.
From a technological point of view, a
multicast group is beneficial for the network, provided that it has 
at least $n^{(l)}$ members. This is due to the signaling overhead of the 
multicast, which depends on the underlying network technology. 

The operator constructs the multicast bids $\mathcal{M}_k^{(l)}$ that
complement the unicast bids $\mathcal{B}_k^{(l)}$ in the auction
by grouping together at least $n^{(l)}$ users requesting the same 
service, e.g. watching a video at a certain quality. Let $g \in \mathcal{G}$ be a 
multicast group. 
The multicast bid is straightforwardly defined as $(p_g, m_g)$,  
where $p_g = \sum_{i \in g}{p_i}$ and $m_g = m_i$.

{\bf Proposition 2:} It is beneficial for the network and harmless for the 
users of a multicast group to have their unicast bids deleted.
\\
\noindent{\bf Proof:\,}
Since $n^{(l)} > 1$, and $p_g = \sum_{i \in g}{p_i} > p_i, \   \forall i \in g$,
a multicast bid always tops its members' unicast bids, thus having 
{\em strictly} higher probability of winning.
Also, since $m_g = m_i < \sum_{i \in g}{m_i}, \   \forall i \in g$,
it is always socially efficient to serve users by means of multicast when
the $n^{(l)}$ constraint is met,
since more users can be served by the network.
\\

Due to Proposition 2, auction winner determination is performed as follows:
The operator deletes the ``redundant'' unicast bids of the users belonging
to some multicast group. He then mergesorts the unicast and multicast bids and 
declares the {\em largest} set of the {\em highest} bids  
that do not violate the capacity constraint as winning. Therefore, the
same algorithm of Section~\ref{sec:auc} is applied to declare the winning
unicast and multicast group bids, denoted as $\mathcal{W}_k^{(l)}$ and 
$\mathcal{WM}_k^{(l)}$ respectively.

{\bf Proposition 3:} Social welfare is maximized.
\\
\noindent{\bf Proof:\,} 
Efficiency is attained, since users bid truthfully, 
i.e. $w_i = u_i$, and after the algorithm terminates it is impossible for
a winning bid to be lower than any kind of losing bid:
\begin{eqnarray}
b_w > b_i &\forall b_w \in \mathcal{W}_k^{(l)},  &\forall b_i \notin \mathcal{W}_k^{(l)}\\
b_g > b_i &\forall b_g \in \mathcal{WM}_k^{(l)}, &\forall b_i \notin \mathcal{WM}_k^{(l)}\\
b_w > b_g &\forall b_w \in \mathcal{W}_k^{(l)},  &\forall b_g \notin \mathcal{WM}_k^{(l)}\\
b_g > b_i &\forall b_g \in \mathcal{WM}_k^{(l)}, &\forall b_i \notin \mathcal{W}_k^{(l)}
\end{eqnarray}
\\

\section{Time, User Utility and Incentives}\label{sec:time}

Defining the user utility for receiving service in
a certain time interval is non-trivial, especially for
long-lived services where the consistent reservation of resources in 
subsequent auctions is highly beneficial. This is for instance the case
for a long-lived real-time streaming video service; clearly losing at some 
auction will result in loss of content and dis-satisfaction for the user
and thus to a reduced willingness to pay for the entire service.
Hence, user utility may depend on the {\em history} of resource allocations. 
History-dependent utility functions capable of expressing such preferences
have been have originally been 
proposed in \cite{ATHENA} for auction-based resource allocation in UMTS networks 
and subsequently used elsewhere \cite{HERA, Bruno}. 
The main merit of history-dependent utility functions is
that {\em multiple} quality parameters such as the vector of instantaneous
bit rates, delay and/or total quantity of resources
allocated impact the values of the {\em correlated marginal utilities}
and the overall expected level of users' satisfaction.
We use the term ``marginal utility'' to denote the 
additional utility attained over each slot of the user's service session.
Thus, these utility functions can accurately quantify the
time-varying user-perceived quality.
These could complement
our scheme by using them to compute the value of $w_i$ 
to be submitted at time $t$, as a function of the user's utility
for the long-lived service and the history of user's allocations so far.

Note also that a small value for the length of time for which the 
auction allocations apply, allows our scheme to quickly
adapt to the varying demand. However, it also implies that 
the auction is run more often. Thus, this value
should be large enough for the auction to run between two consecutive allocation 
intervals. This obviously depends on the complexity of the auction,
which is derived in Section~\ref{sec:assessment}.

\begin{figure}[htbp]
	\centering\includegraphics[width=.6\columnwidth]{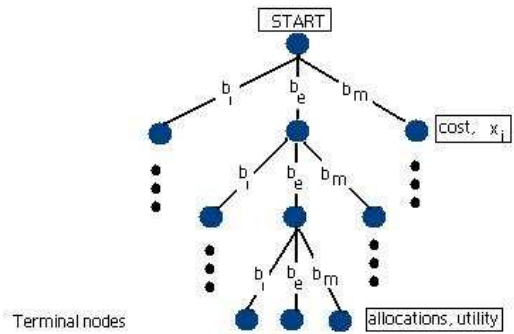}
	\caption{The sequential form of the repeated game for 3 auctions.}
	\label{fig:sgame}
\end{figure}

We now briefly address the issue of user incentives for this repeated game,
depicted as Fig.~\ref{fig:sgame}, where users bid in a sequence of auctions. 
Node START denotes user's $i$ start of bidding and in general each node 
({\em state}) corresponds to the bidding phase of each auction he participates. 
At each node user $i$ selects an {\em action}, i.e. to bid truthfully
$b_e = w_i$, or shade his bid $b_l < w_i$, or bid aggressively $b_m > w_i$.
We denote the corresponding bidding strategies as $S_e$, $S_l$ and $S_m$
respectively.

Incentive compatibility still holds if user bids for independent services at the
various auctions. Since the user utility is additive, 
so is the expected payoff from the  game, thus $EP = \sum_{t=1}^{T} EP_t$. 
In order to maximize this sum, it suffices to maximize all the $EP_t$. However, 
$EP_t = Pr(win) \cdot (u_j - SocialOpportunityCost)$. Since the $SocialOpportunityCost$
is unaffected by customer's own bids and the probability of winning is
maximized if $b_j = u_j$, so are $EP_t$ and $EP$. This means that incentive
compatibility is bidders' dominant strategy for the aforementioned game.
Thus, since it is best for bidder always to follow strategy $S_e$, the realization 
probabilities vector for the strategies $<S_l, S_e, S_m>$ is $<0, 1, 0>$. Note that
this is not just a Nash equilibrium strategy, it is dominant strategy and
{\em subgame perfect} (always best to reveal with his bid his true valuation, regardless
of the node of the tree of the game where he is located).

We now address the most interesting case where {\em complementarities} exist 
among the user allocations in subsequent auctions. In this case, wining at
one of a series of auctions brings in addition to the value $u_i$
an extra net benefit by increasing the value of future allocations; this is 
because losing at some auction for a user e.g. of a video service would result
in lower value for the whole of the service due to the incurred service interrupt. 
Thus, the question is whether strategy $S_m$ 
or a mixed strategy comprising of the strategies $S_e$ and $S_m$,
could result in higher expected payoff than $S_e$. This 
might be indeed the case if the bid  $b_e$ is at some auction slightly lower 
than the cutoff price and a bid $b_m$ 
brings extra value (due to service continuation) higher than the extra 
charge paid, that he would lose if bidding $b_e$. 
However, this would {\em not} be the case for 
extremely ``uncertainty averse'' (conservative) 
users, who - by definition - in cases of choice/behavior under  {uncertainty} 
always opt to play the safest strategy. Therefore, for this type of users
truthful bidding comprises a subgame perfect equilibrium strategy.  
This ``maximin'' behavior was  proposed by Wald \cite{Wald} for situations of 
severe uncertainty, which is also encountered by the bidders in our auction. 

\section{Cooperative Schemes}\label{sec:coop}

In this section we show how the proposed mechanism can be transformed to
a cooperative bandwidth allocation mechanism for
operators who prefer flat rate pricing to usage based pricing schemes. 
In this context, the user utility for the service $w_i \forall i$ is replaced
by a predefined weight $w_s$ that the operator assigns to each {\em type} of
service $s$. This modification suffices to modify the auction to a 
cooperative bandwidth allocation scheme. Note that under this
modification, determining the payments of the winners 
is performed instantly, i.e. in $O(1)$.

Due to the different weights assigned per service type, 
this scheme prioritizes the services having greater weights. 
These services will enjoy statistically higher quality than others
and this scheme serves essentially as a DiffServ mechanism.

Also, the operator can assign different weights per time of day
in the various services. This way, in peak hours he may discourage  
demanding services by assigning 
a low weight, and also emulate time-of-day pricing. 

Furthermore, weights can be dynamically computed from weighting functions per
flow, so that each weight takes into account the flow's
overall service time to prioritize older flows, emulating
schemes like CHiPS \cite{chips} where the winner of an auction is
prioritized compared with new flows.

Last but not least, it is also possible to modify the mechanism so that it
can perform well-known scheduling policies such as Round Robin and First
Come First Served. Indeed, in order to do so it suffices to define each flow's weight 
to be equal to the inverse of the assigned share of resources and the inverse 
of the time of the flow initiation respectively. Obviously, these scheduling
policies are not economic-aware and thus it is impossible for them to outperform
the auction in terms of the attained social welfare.

This comprises further evidence that the proposed mechanism is flexible enough to
be tailored and customized to various resource allocation policies that the 
network operator may wish to employ. This is also an attractive feature of our
mechanism since it can be envisioned that in a Wireless Future Internet network
architecture, it would be both possible and desirable for a provider to be able 
to employ different such policies over a certain part of his network capacity, thus
applying both richer and more sophisticated resource allocation policies.

\section{Assessment}\label{sec:assessment}

Having assessed our mechanism in economic terms, by taking into advantage 
the fact that our mechanism is essentially a VCG mechanism, we proceed to 
assess its computational complexity.

\subsection{Auction Complexity}
Sorting the bids of each link auction is done in $O(N \cdot logN)$, with $N$ denoting 
the total number of bids. Winner determination is then done in $O(N)$, so that the
point where the capacity constraint is violated is found. 
Computing winners' charge exploits the fact that user rates are not arbitrary but
pertain to discrete service rates: First, we compute the charge for all
service rates by adding the highest losing bids whose sum of rates equals this rate. 
Then each winner is charged his respective service charge. 
This
is bounded by $O(s \cdot N)$, with $s$ denoting the number of different service rates.
Deletion of bids can be done in $O(N \cdot logN)$ since finding each bid can be done in 
$O(logN)$ using binary search and this must be performed for at most $N$ bids.
Since same-tier auctions run in parallel, while different tier auctions sequentially, 
the mechanism's overall complexity is bounded by 
$O(L \cdot (N + s \cdot N  + N \cdot logN))$.

\subsection{Multicast Extension Complexity}
The additional overhead of the multicast extension is due to creating
multicast bids and deleting redundant unicast bids. The former can be
performed 
by parsing {\em once} the sorted list of bids and
classify the users to separate groups, based on their selection of service,
for which the total numbers of users and willingness to pay are updated 
as the list is constructed. This is done in $O(N)$, with $N$ denoting the 
total number of bids. Subsequently, each redundant unicast bid must be 
deleted. Finding each bid can be done in $O(logN)$ using binary search. 
Since at most $N$ bids must be deleted, the complexity bound of 
multicast is $O(N + N \cdot logN)$.
 
\section{Simulations}\label{sec:sims}

In this section we assess experimentally various aspects of the 
proposed auction  by means of simulations. We have implemented
the proposed mechanism as a Java application. 
We have run numerous simulation experiments according to 
a detailed simulation model, specifying the distributions of 
user arrivals, departures, and service requests, and the mix
of users in terms of the number of users per service requested and the
distribution of their total willingness to pay.
For each user, the total willingness to pay is randomly selected according 
to a uniform distribution over an interval which is determined by 
whether the user is of low, medium or high value. 
The simulation is run for a series of $T$ auctions where a
number of users bid; their service start time $t_s$ is drawn from 
a uniform distribution having support in $[1, T]$ and duration
are also drawn uniformly from $[1, T-t_s]$.
The total quantity of resource units available at 
each auction also fluctuates (due to the varying 
resources available) in the simulation model, 
and is randomly selected according to a uniform distribution. 
The capacities for the networks used in the simulations run typically 
match that of standard network interfaces, i.e. WiMax, 3G HSPA and 
WLAN networks. Finally, the user services used for the simulations
are CBR streaming video of low quality (Video-LQ) of 1 Mbps, video of
high quality (Video-HQ) of 5 Mbps and FTP of 1 Mbps.

For brevity reasons, rather than describing in detail the various simulations
conducted, we present the main findings and provide some illustrative examples.

\subsection{Service completion and handoffs}

Adopting an auction mechanism comprising of a series of auctions for resource
allocation in consecutive time periods may affect considerably the resource
allocation pattern for a long-lived service. For instance, if a user is forced
to participate in 50 auctions in order to receive a video service, it is possible
that the service is interrupted at some auctions if the user's bid is not
high enough. This is inevitable due to the varying competition, however it is
desirable that this does not occur often. Indeed, if it is common for users 
to experience service interrupts, they would be displeased and the applicability
of the proposed auction would be limited. In fact, this is a typical argument
in favor of non economic-aware scheduling policies such as FCFS that do not
exhibit this problem.

We have run various simulations in order to assess the percentage of the
total resources that users receive when bidding in a series of auctions.
Our simulations indicate that due to the hierarchical form of the network
and its large multiplexing capacity, the proposed auction typically serves
users in a satisfactory way. Indeed, the percentage of users that receive 
mediocre service is limited, while the vast majority of users is either 
served perfectly or not served at all. 

\begin{figure}[htbp]
	\centering\includegraphics[width=.6\columnwidth]{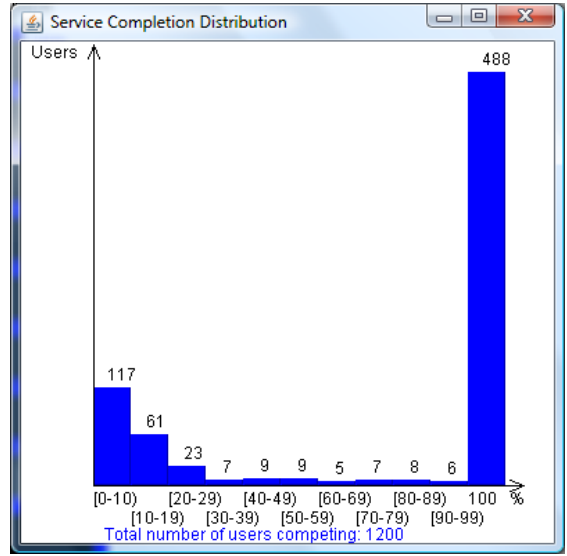}
	\caption{Service completion percentages for the auction bidders.}
	\label{fig:service}
\end{figure}

We provide as Fig.~\ref{fig:service}  the percentage of users
service completion for  a series of 500 auctions and a network comprising of
1 wide area network of 500 Mbps, 2 medium area networks of 50 Mbps and
10 local area networks of 6 Mbps, each of which serves 40 FTP users with
medium willingness to pay, 40 Video-LQ users with medium willingness to pay and
40 Video-HQ users with medium willingness to pay. Note that due to the fact
that the willingness to pay of the users is drawn from the same interval
of uniform distribution and due to the dynamic arrivals and departures over
time, this is the least favorable scenario for the proposed auction, since 
users with higher willingness to pay are likely to arrive and displace some
customers that already receive service. However, the auction performs very
well. Note that in order to make the plot more readable, we plot the percentages
of the 740 users that won at least in one auction; the remaining 460 of the total
1200 users had bids that were always under the auction cut-off price and were
never served.

It is also worth emphasizing that this performance may be further
improved by adopting a scheme that prioritizes users that have already won 
at past auctions as opposed to new arrivals, such as CHiPS \cite{chips};
simulating and assessing such schemes in the context of our auction comprises
an interesting direction of future research.

\begin{figure}[htbp]
	\centering\includegraphics[width=.6\columnwidth]{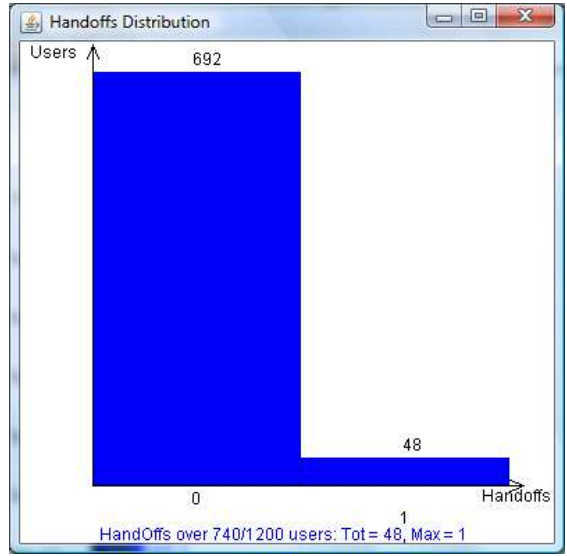}
	\caption{Number of handoffs experienced by the users.}
	\label{fig:handoff}
\end{figure}

Finally, we comment on the number of handoffs users experienced throughout
their service time. In the simulation set up described above, users never
experienced more than one handoff throughout their service time, as 
depicted also in Fig.~\ref{fig:handoff}. A handoff
is defined in our context as a change in the serving network interface of
a user in two consecutive auctions within his service time. Note that we
do not measure as handoff a switch in the serving network after a service
interrupt, though this typically happens in the simulations run, 
since this does not result in additional complexity for the
network. From the simulations run, it is concluded that the maximum number
of handoffs depends heavily on the relation of the network interfaces - and 
especially those of the lowest geographical coverage - capacity
and the rate of the user services; the higher the multiplexing capacity
of the networks, the higher the number of handoffs experienced.

\subsection{Speed of execution}

We have conducted many simulations in order to assess the major factors 
that affect the speed of the auction execution. These simulations indicate
that the dominant factor that determines the speed of execution is the number
of end customers participating. The way these customers are distributed or the
number of networks where these customers can be assigned to have a second order
impact compared to the total number of bidders. The impact of the capacity of
the networks comprising the hierarchical network access is negligible.

In particular, one set of simulations conducted comprises of 100 simulations 
for a 3-tier network where
there is one wide-range network of capacity 500 Mbps; within its coverage
there are two networks of capacity 50 Mbps and 4 networks of 10 Mbps.
The auction was run and its execution time was measured in a laptop with 
AMD Turion TL50 processor with 1 GB DDR2 running Windows XP. The average
execution time of $T=100$ consecutive auctions and 120 users was 2374 msec. 
The measured execution time was 15485 msec for 600 users and 39487 msec 
for 1200 users. 

This example depicts both the impact of the number
of users in the total execution time but also the limited time required to
run the auction in practice: the execution of one instance of the auction,
which would be the case in a real network, can be typically done in an order
of magnitude of msecs or few seconds. This is further evidence of the proposed
auction's applicability in practice.

\subsection{Revenue and social welfare}

It is no surprise that the auction revenue and social welfare depends on 
the relationship between demand and supply. This is true for all auction
mechanism and also applies to the proposed mechanism as well. Indeed, 
in the simulation conducted it has been observed that the most fierce the 
competition, the higher the attained revenue and the closer it is to the
social welfare attained. Though the revenue and social welfare values should
be computed for the entire ``hierarchical'' auction, we provide some indicative
plots. These plots indicate that the revenue and social welfare values among 
networks of the same tier are quite similar. This is due to the uniform way 
we generate demand and distribute it to the end tier networks. Also, due to
the way the auction is run and the higher capacity of the higher tier networks
both the social welfare and revenue values for these interfaces are considerably
higher. 

\begin{figure}[htbp]
	\centering\includegraphics[width=.6\columnwidth]{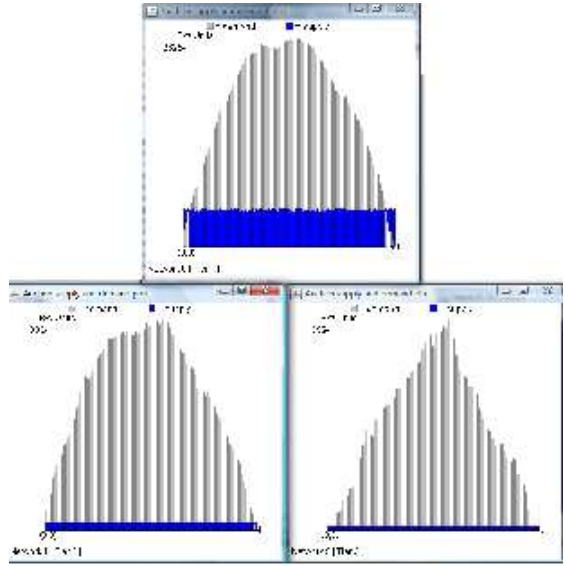}
	\caption{Demand and supply distribution among the network interfaces.}
	\label{fig:demand}
\end{figure}

\begin{figure}[htbp]
	\centering\includegraphics[width=.6\columnwidth]{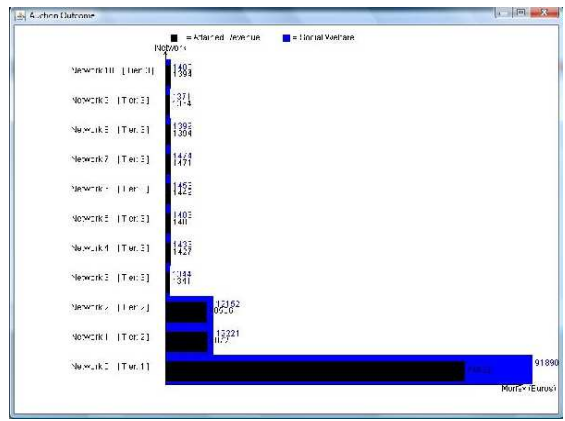}
	\caption{Revenue and Social Welfare plots.}
	\label{fig:revSW}
\end{figure}

Below, we provide some indicative plots, namely Fig.~\ref{fig:demand}
and Fig.~\ref{fig:revSW}, that depict the distribution of demand,
supply, social welfare and attained revenue for a series of 100 auctions for 
a hierarchical network comprising of
1 wide area network of 500 Mbps, 2 medium area networks of 50 Mbps and
8 local area networks of 6 Mbps, each of which serves 50 FTP users with
high willingness to pay, 50 Video-LQ users with low willingness to pay and
50 Video-HQ users with medium willingness to pay.

\section{Conclusions}\label{sec:concl}

In this paper we have presented a mechanism for the allocation of the downlink 
bandwidth of a Wireless Future Internet network, whose access architecture is 
hierarchical in terms of radius and thus geographical coverage.  
In particular, we have designed an incentive-compatible auction mechanism 
of low computational complexity. We have extended the mechanism so that 
multicast is supported, defined a repeated game to study utility and incentives
issues and transformed our auction to a cooperative 
scheme for services prioritization. We have assessed the proposed mechanism
in economic, game-theoretic and complexity terms and its effectiveness by 
means of simulations.
Both the theoretical and the experimental analysis indicate that the mechanism 
specified in this paper is  efficient, fast and an attractive means of 
economic-aware resource allocation. 
Defining weighting functions for 
emulating DiffServ and CHiPS with
our cooperative mechanism comprises interesting direction of future research.

\end{document}